\documentclass{article}

\usepackage{arxiv}

\usepackage[utf8]{inputenc} 
\usepackage[T1]{fontenc}    
\usepackage{hyperref}       
\usepackage{url}            
\usepackage{booktabs}       
\usepackage{amsfonts}       
\usepackage{nicefrac}       
\usepackage{microtype}      
\usepackage{lipsum}
\usepackage{graphicx}

\title{STM/S Observations of Graphene on SiC(0001) Etched by H-plasma}

\author{
  Andr\'e E. B. Amend$^{1}$\thanks{E-mail: amend@kelvin.phys.s.u-tokyo.ac.jp}, Tomohiro Matsui$^{1}$\thanks{E-mail: matsui@phys.s.u-tokyo.ac.jp}, Hiroki Hibino$^{2,3}$ and Hiroshi Fukuyama$^{1,4}$\\
$^{1}$Department of Physics, The University of Tokyo, 7-3-1 Hongo, Bunkyo-ku, Tokyo 113-0033, Japan \\
$^{2}$School of Science and Technology, Kwansei Gakuin University, 2-1 Gakuen, Sanda, Hyogo 669-1337, Japan \\ 
$^{3}$NTT Basic Research Laboratories, 3-1 Morinosato Wakamiya, Atsugi, Kanagawa 243-0198, Japan \\
$^{4}$Cryogenic Research Center, The University of Tokyo, 2-11-16 Yayoi, Bunkyo-ku, Tokyo 113-0032, Japan\\
}

\begin{document}
\maketitle

\begin{abstract}
Monolayer graphene epitaxially grown on SiC(0001) was etched by H-plasma and studied by scanning tunneling microscopy and spectroscopy. The etching created partly hexagonal nanopits of monatomic depth as well as elevated regions with a height of about 0.12~nm which are stable at $T = 78$~K. The symmetric tunnel spectrum about the Femi energy and the absence of a $6\times6$ corrugation on the elevated regions suggest that in these regions the carbon buffer layer is decoupled from the SiC substrate and quasi-free-standing bilayer graphene appears at originally monolayer graphene on the buffer layer. This is a result of passivation of the SiC substrate by intercalated hydrogen as in previous reports for graphene on SiC(0001) heat treated in atomic hydrogen.
\end{abstract}

\section{Introduction}
The crystalline monatomic carbon sheet, graphene, has been in the spotlight of research for many years due to its stability at atmospheric conditions despite ultimate thinness, and its remarkable electronic and mechanical properties~\cite{Neto2009,Abergel2010,Review3}. One of the peculiar features is the electronic structure of graphene edges, which strongly depends on their atomic configuration. Zigzag edges, which feature only one of the two graphene sublattices at the edge, are of special interest because they host degenerate electronic flat-bands at the charge neutrality point ($E_{\mathrm{D}}$) that appear as the edge states~\cite{Fujita1996,Niimi2005,Kobayashi2005,Niimi2006}. As long as the flat bands are located near the Fermi energy ($E_{\mathrm F}$), it becomes energetically favorable for electronic spins along the edges to align. The magnetic polarization is, however, not stable at isolated edges. Instead, at two zigzag edges on a nanoribbon narrower than $\sim$10~nm, the edge spin alignment is theoretically stable even for small electron interaction~\cite{Fujita1996,Son2006,Jiang2008}. Several experimental studies~\cite{Tao2011,Baringhaus2013,Li2014,Magda2014,Ruffieux2016,Wang2016} have claimed observations of the spin-polarized edge state in narrow zigzag nanoribbons (zGNR), but it still remains controversial. This is partly because a lack of atomic precision of the edge shape, edge termination or strong doping by the substrate made polarization unlikely, or because ribbon dimensions were limited by the experimental methods.

To fabricate narrow and precise zGNRs on which spin-polarization could be verified, anisotropic hydrogen (H) plasma etching is a promising method. For graphite, it etches a single graphene layer on the surface and produces well-shaped hexagonal nanopits with expectedly hydrogen-terminated zigzag edges~\cite{Yang2010,Amend2018,MatsuiTBP}. However, nanoribbons fabricated on graphite interact with the underlying graphite substrate, which is not considered in most theoretical calculations of the spin-polarized edge state. 
In pristine epitaxial few-layer graphene on SiC(0001), spin polarization is considered not to be stable because of relatively strong electron-doping by the substrate. On the other hand, in epitaxial graphene on SiC(0001), quasi-free-standing graphene (QFSG) with minimal substrate influence has been obtained through intercalation by atomic hydrogen~\cite{Riedl2009,Watcharinyanon2011,Tanabe2012,Murata2013,Murata2018}. The problem of the doping can be reduced in the QFSG after decoupling from the substrate~\cite{Riedl2009}. Thus nanoribbons fabricated at QFSG on SiC(0001) by H-plasma etching is a hopeful material to study the spin-polarized edge states. Very recently, H-plasma etching of graphene on SiC(0001) with initial defects has been investigated by Wu \textit{et al.}~\cite{Wu2018} with scanning tunneling microscopy (STM) and spectroscopy (STS), where nanopits aligned in zigzag direction are successfully created. However, in their work, no QFSG was observed and tunnel spectra obtained under a strong buffer layer influence were reported.

In this work, we attempted to realize both the hexagonal nanopit and QFSG formations by exposing graphene on SiC(0001) to H-plasma at a temperature of 600$^{\circ}$C. It was verified, through STM/S measurements at low temperature, that both the processes occur simultaneously. Epitaxial graphene samples prepared by H-plasma etching thus potentially yield zigzag nanoribbons on both the buffer layer and on QFSG, which makes this system useful to study the substrate influence on the electronic state of such nanoribbons with zigzag edges.

\section{Experimental}
For this study, mono-and bilayer graphene samples were epitaxially grown on a 4H-SiC(0001) surface. The number of layers was estimated from the phase contrast in atomic force microscopy images. For etching, H-plasma is generated by applying a radio frequency power ($W_{\mathrm{RF}}=$~20~W, $f_{\mathrm{RF}}=13.56$~MHz) to a coil through which hydrogen gas flows upstream of the sample. The hydrogen gas pressure is controlled to be $P = 110$~Pa at the sample location. The sample is heated to a temperature of 600$^\circ$C and exposed to the plasma for 10 min. A graphite sample was etched together with the epitaxial graphene for comparison. More details of the etching apparatus and technique will be described elsewhere~\cite{MatsuiTBP}.

STM/S measurements were conducted in an ultra-high vacuum ($\le10^{-8}$ Pa) at $T=$~78~K to analyze the etched graphene on SiC(0001). The surface topography as well as local spectra of differential tunnel conductance (d$I$/d$V$), which is proportional to the local density of states, were measured. The graphite sample was studied by STM at room temperature in air for comparison.

\section{Results and Discussion}
\begin{figure}[b] 
\centering
\includegraphics[width=0.6\textwidth,keepaspectratio]{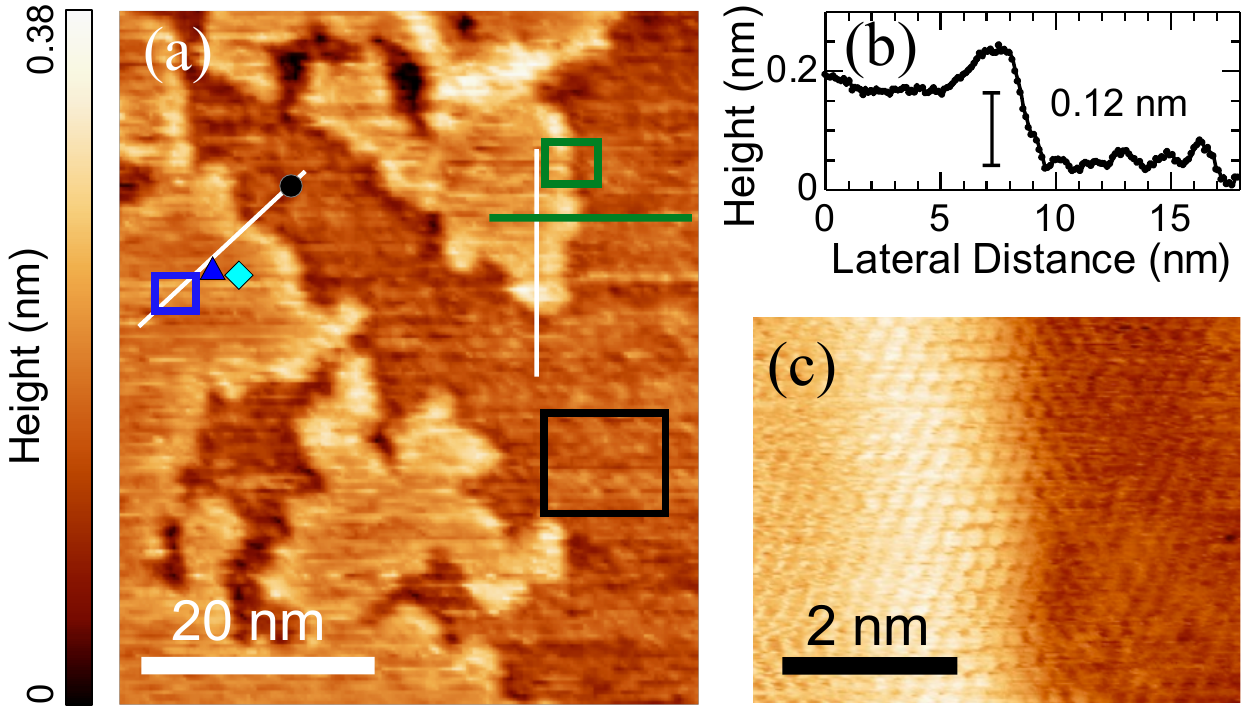}
\caption{(a) Topographic STM image showing elevated terraces on epitaxial graphene on SiC(0001) after the H-plasma treatment. (b) Height profile across the island boundary along the green line in (a). (c) STM image showing continuous atomic rows across the boundary in the the green rectangle in (a). The symbols and the other rectangles in (a) refer to the measurements shown in Fig. \ref{f2}, the white lines in (a) to those in Fig. \ref{f3}. The STM measurement parameters are $V_{\mathrm{bias}}=$~(a)~500~mV,~(b,c)~250~mV, and $I_{\mathrm{t}}=$~(a)~0.05~nA,~(b,c)~0.06~nA.}
\label{f1}
\end{figure}
The surface morphology is changed drastically by the H-plasma etching, becoming covered by many nanopits and elevated terraces with complicatedly shaped boundaries on more than half of its surface. Figure~\ref{f1}(a) shows such a boundary area, where brighter regions correspond to the elevated surface and darker regions to non-elevated areas. The line profile (fig.~\ref{f1}(b)) across one of the boundaries along the green line indicated in (a) shows that the brighter area is elevated by 0.12~$\pm$~0.03~nm, which is much smaller than the inter-layer distance of graphite (0.33~nm) or a SiC step (0.25~nm). The high resolution image (fig.~\ref{f1}(c)) taken in the green rectangle in (a) reveals an atomic corrugation of carbon atoms that continues across the boundary. This indicates that the same graphene sheet covers both terraces. 

\begin{figure}[h] 
\centering
\includegraphics[width=0.6\textwidth,keepaspectratio]{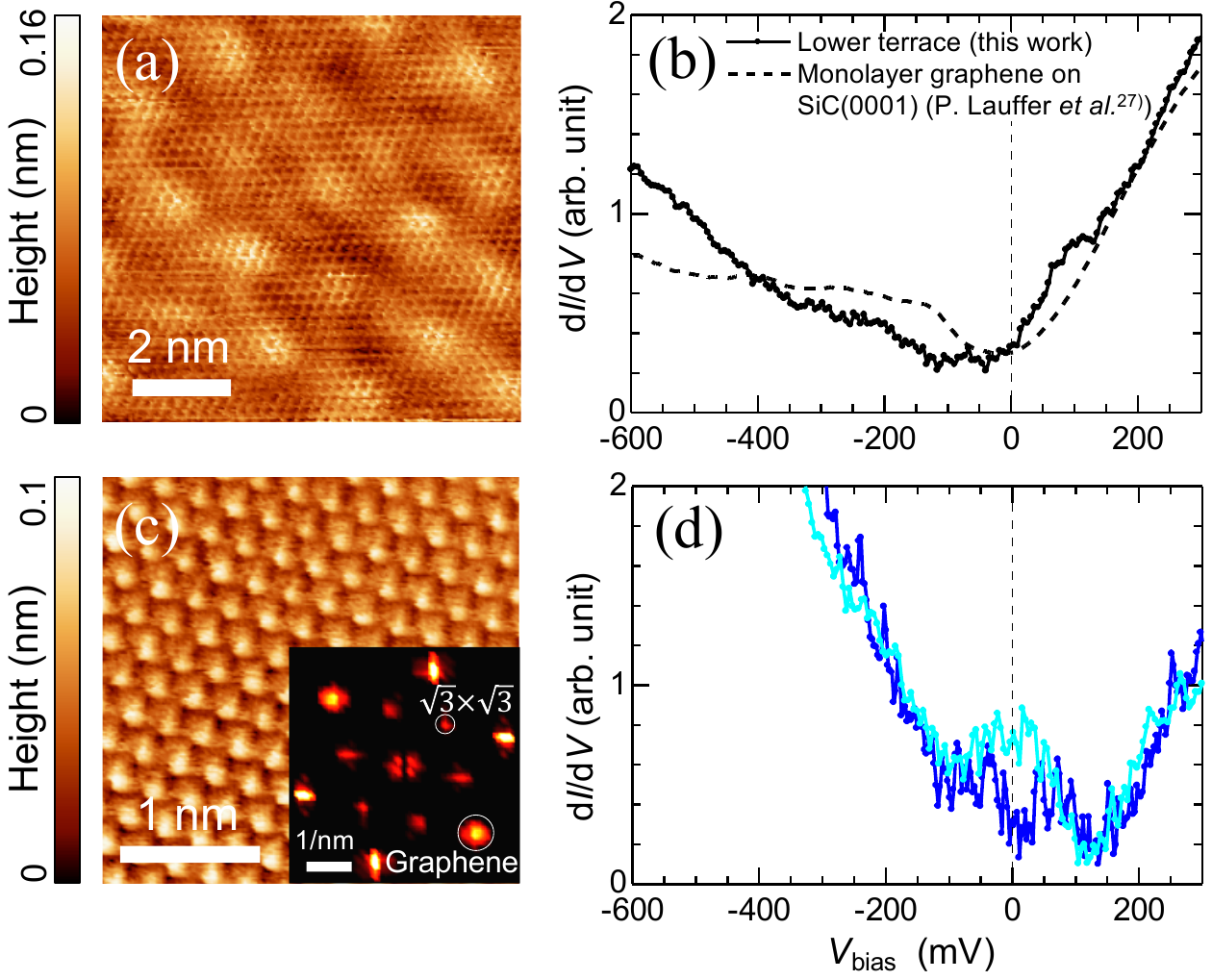} 
\vspace{-1em}
\caption{(a) STM image and (b) d$I$/d$V$ spectrum on the lower terrace 
taken in the black rectangle and at the black dot indicated in fig.~\ref{f1}(a), respectively. The dashed line in (b) is the spectrum of epitaxial graphene on SiC(0001) published by Lauffer \textit{et al.} \cite{Lauffer2008}. (c) STM image and (d) d$I$/d$V$ spectra on the elevated terrace taken in the blue rectangle and at the blue triangle and diamond indicated in fig.~\ref{f1}(a), respectively. The inset in (c) shows the Fourier transform of the STM image in the main figure. Here, the bright spots of the triangular lattice and $\sqrt{3}\times\sqrt{3}$ superlattice are marked. The STM and d$I$/d$V$ measurement parameters are $V_{\mathrm{bias}}=$~(a)~500~mV, (c)~300~mV, $I_{\mathrm{t}}=$~(a)~0.05~nA, (c)~0.06~nA, and $V_{\mathrm{mod}}=$~(b,d)~3~mV.}
\label{f2}
\end{figure}
Figure \ref{f2}(a) shows an STM image for the location indicated by the black rectangle in fig.~\ref{f1}(a) corresponding to a lower terrace. On this terrace, a $6\times6$ corrugation is observed together with the atomic corrugation. This superlattice corrugation typically is induced on few-layer epitaxial graphene by the $6\sqrt{3}\times6\sqrt{3}$ corrugation of the C-rich buffer layer~\cite{Kaplan1989,Northrup1995}. Its presence thus signifies that the graphene is supported by the buffer layer. The d$I$/d$V$ spectrum (solid line in fig. \ref{f2}(b), taken at black dot in fig. \ref{f1}(a)) observed on the lower terrace is consistent with this inference. The spectral weight below $E_\mathrm{F}$ is substantially suppressed, because $E_{\mathrm{D}}$ is located on the negative energy side, resulting in a highly asymmetric spectrum with respect to $E_\mathrm{F}$. This characteristic feature is very similar to the well-known spectrum of monolayer graphene on SiC(0001) measured by Lauffer \textit{et al.}~\cite{Lauffer2008}, shown by the dashed line in fig. \ref{f2}(b). We thus conclude that the lower terrace in this study is the original monolayer graphene on SiC(0001).

In contrast to the lower terrace, only the triangular atomic lattice without any $6\times6$ corrugation was observed on the elevated terrace as shown in Fig.~\ref{f2}(c) (taken in the blue rectangle in fig. \ref{f1}(a)). The absence of the superlattice corrugation suggests that the graphene is not supported by the buffer layer. Because only one sublattice is visible here, it cannot be free-standing \textit{monolayer} graphene. Instead, quasi-free-standing \textit{bilayer} graphene is expected to appear if the former buffer layer were decoupled from the SiC(0001) substrate by intercalating hydrogen that passivates the substrate. This hypothesis is confirmed by the d$I$/d$V$ spectra shown in fig. \ref{f2}(d) (taken at the blue triangle and diamond in fig. \ref{f1}(a)), for which the asymmetric feature of the lower terrace is significantly reduced becoming almost symmetric for $|V_\mathrm{bias}|\geq100$~mV on the elevated terrace. Note that it does not reach zero at $|V_\mathrm{bias}|\leq100$~mV, but stays finite and hosts peaks. The peak energies depend on the measured area, but typically remain constant within more than $10$~nm. Nevertheless, the symmetric d$I$/d$V$ implies that the $E_{\mathrm{D}}$ is located around $E_\mathrm{F}$, and that the bilayer graphene is mostly decoupled from the substrate in the elevated region.

It is known that intercalating atomic hydrogen decouples the buffer layer at high temperatures, thus changing monolayer graphene on a buffer layer to bilayer QFSG~\cite{Riedl2009, Watcharinyanon2011,Tanabe2012}. Noting that the elevated terrace height is consistent with that in published data\cite{Watcharinyanon2011,Murata2013}, and that $E_\mathrm{D}$ is located near $E_\mathrm{F}$, we suspect that a similar decoupling by H-intercalation is realized by the H-plasma treatment providing bilayer QFSG. The position-dependent low-energy features in Fig.~\ref{f2}(d), as well as the $\sqrt{3}\times\sqrt{3}$ superlattice, that can be seen in the Fourier transform (inset of fig.~\ref{f2}(c)), suggest inhomogeneous H-intercalation with a finite concentration of defects or dangling bonds. Those defects should be responsible for the low energy peaks in the tunnel spectra near $E_{\mathrm{F}}$ (fig.~\ref{f2}(d)). It is reported in $\it{monolayer}$ QFSG prepared by atomic hydrogen~\cite{Murata2018} that the system is heavily hole doped and that there appear many nanometer-size bright spots in the STM images consisting of clusters of several dangling bonds. In our $\it{bilayer}$ sample prepared by H-plasma, the doping is much lighter and structures related to the clustering of dangling-bonds are not observed.

\begin{figure}[t] 
\vspace{-0em}
\centering
\includegraphics[width=0.9\textwidth,height=\textheight,keepaspectratio]{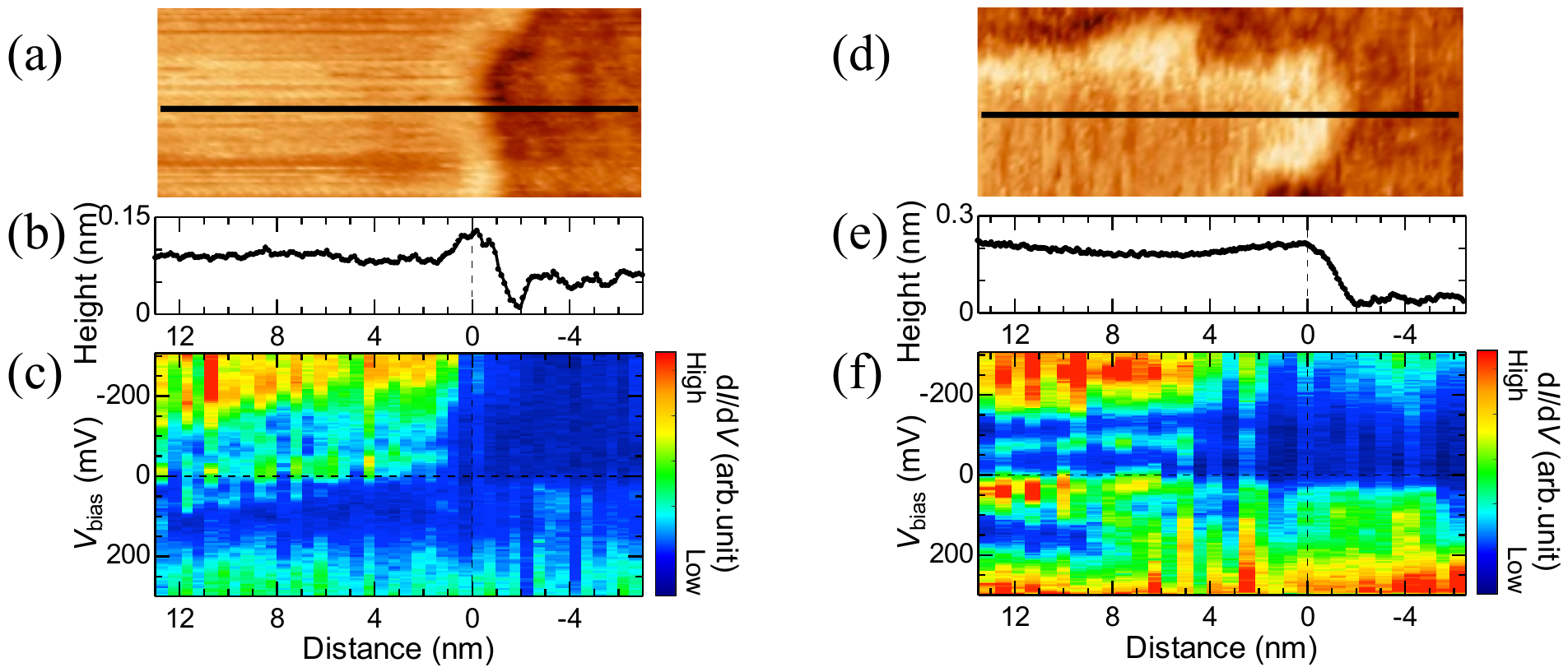}
\vspace{-1em}
\caption{(a) STM image, (b) height profile, and (c) d$I$/d$V$ colormap across the boundary between the elevated and lower terraces along the diagonal white line indicated in fig.~\ref{f1}(a). These are typical data across the boundary. (d) STM image, (e) height profile, and (f) d$I$/d$V$ colormap across the boundary with the peninsula-like structure along the vertical white line indicated in fig.~\ref{f1}(a). The STM and d$I$/d$V$ measurement parameters are $V_{\mathrm{bias}}=$~(a,d,e)~300~mV,~(b)~500~mV, $I_{\mathrm{t}}=$~(a,c,d,e)~0.06~nA, and $V_{\mathrm{mod}}=$~(c,f)~5~mV.}
\label{f3} 
\end{figure}

Figure~\ref{f3}(c) shows a typical spatial variation of the tunnel spectrum plotted as a d$I$/d$V$ colormap along the line indicated in fig.~\ref{f3}(a), a topographic STM image, across the boundary between the elevated ($\it{left}$) and lower terrace ($\it{right}$) regions. Figure~\ref{f3}(b) is the height profile along the same line. The change between the symmetric spectrum together with a peak near $E_{\mathrm{F}}$ on the elevated region and the highly asymmetric one on the lower terrace is clearly seen in the colormap. As expected, the change takes place almost at the same position as the topographic step shown in (a) and (b), at which the distance along the line ($x$) is defined as $x=0$. Such a coincidence between the spectral change and the topographic step does not occur at some boundaries. For example, at the boundary shown in figs.~\ref{f3}(d-f) where the elevated terrace forms a peninsula-like structure, there is an intermediate region at $0 \leq x \leq 3$~nm. In this region, the tunnel spectrum gradually changes from symmetric with low-energy peaks to asymmetric without the peaks, towards the boundary of the lower terrace.

\begin{figure}[h!] 
\vspace{-0em}
\centering
\includegraphics[width=1.0\textwidth,keepaspectratio]{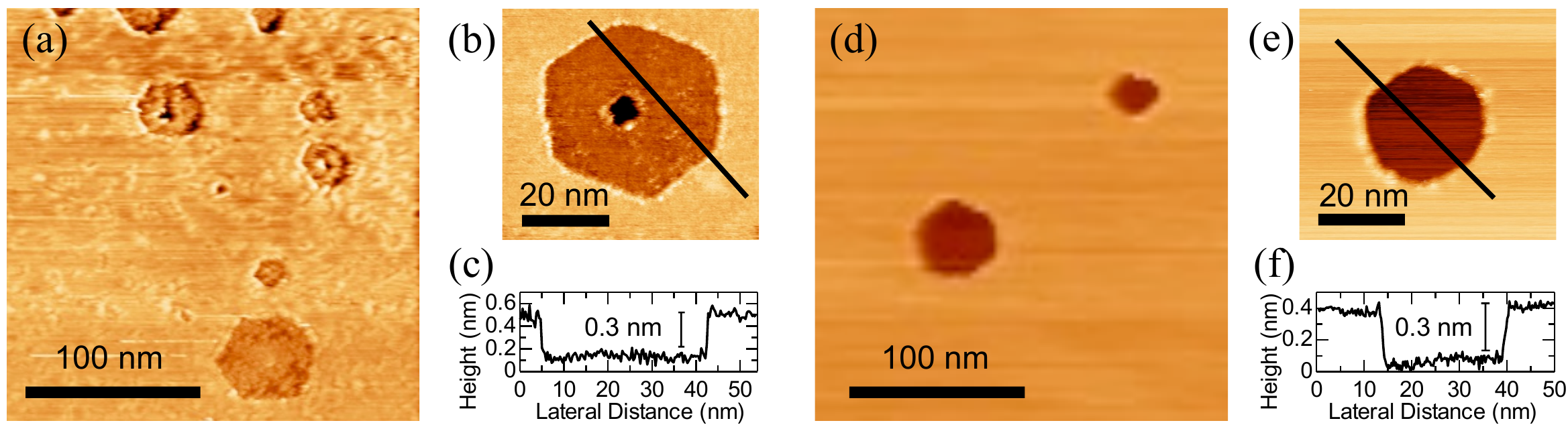}
\caption{(a)(b) STM images and (c) height profile along the line indicated in (b) of hexagonal nanopits created on epitaxial graphene on SiC(0001) by H-plasma etching. (d-f) Similar measurements of hexagonal nanopits created on graphite by H-plasma etching. These two samples were etched simultaneously under the same H-plasma conditions. The nanopits on graphene on SiC(0001) have a higher density than those on graphite, but are irregular in some cases. The STM measurement parameters are $V_{\mathrm{bias}}=$~(a,d-f)~500~mV,~(b,c)~$-400$~mV, $I_{\mathrm{t}}=$~(a-c)~0.06~nA,~(d-f)~0.1~nA.}
\label{f4}
\end{figure}
\begin{figure}[h] 
\centering
\includegraphics[width=1.0\textwidth,keepaspectratio]{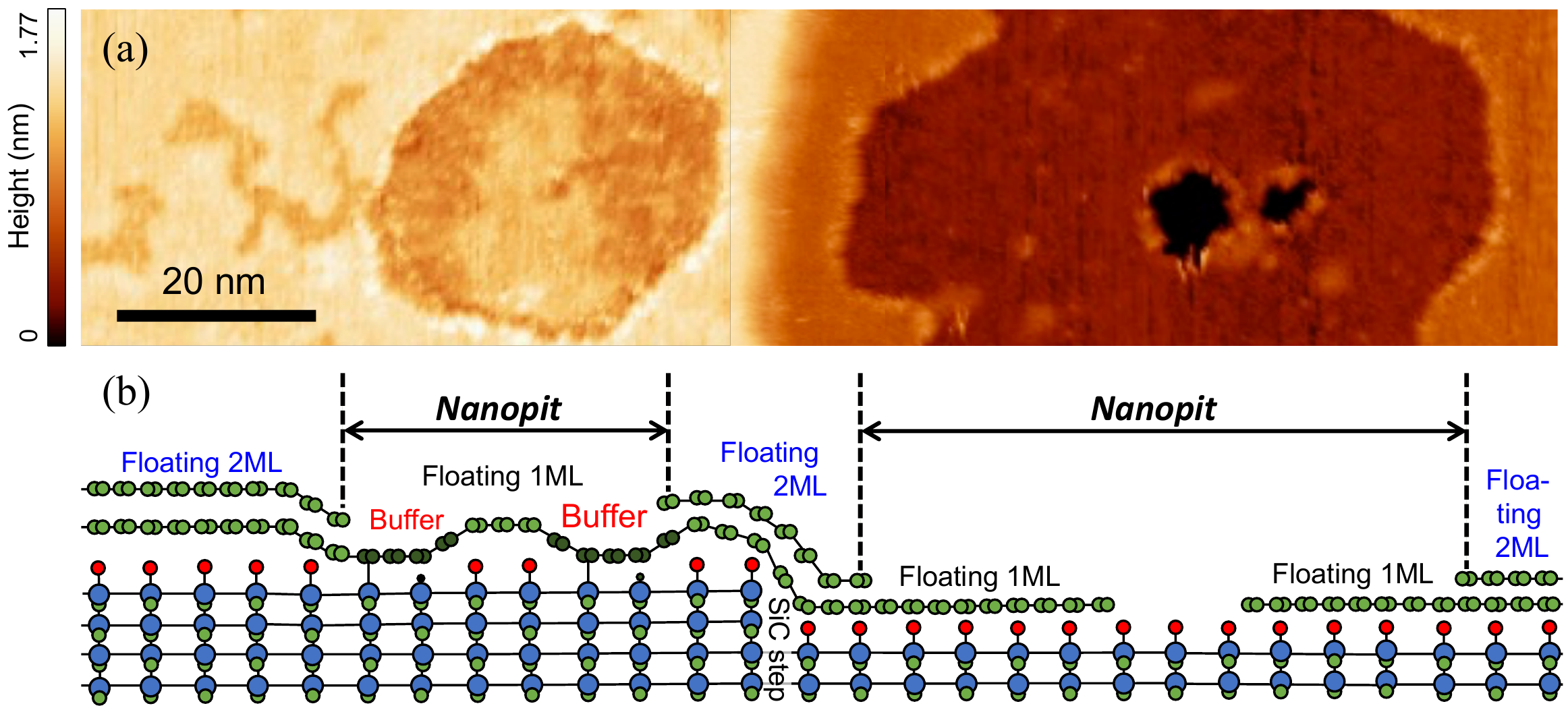}
\caption{
(a) Large scale STM image of a surface of epitaxial graphene on SiC(0001) etched by H-plasma. The measurement parameters are $V_{\mathrm{bias}}=$~500~mV, $I_{\mathrm{t}}=$~0.06~nA. (b) Cartoon explaining the crosssectional structure of the area shown in (a). The green, blue and red circles represent carbon, silicon and hydrogen atoms, respectively.}
\label{f5}
\end{figure}

Together with the elevated terrace we also obtained nanopits by the H-plasma etching, as on the surface of graphite. Figure \ref{f4} shows the nanopits created in graphene on SiC(0001) (a-c) and graphite (d-f) samples that were etched together. The area of the epitaxial graphene in (a) is mostly elevated, but there are small non-elevated regions, which make it rougher than that of graphite in (d). Despite the similarities of the nanopits in both samples, the maximum size of the nanopits is larger and the density of nanopits is higher in epitaxial graphene on SiC(0001) than in graphite. The larger nanopits size of monolayer graphene is consistent with the previous studies~\cite{Yang2010,Diankov2013}. Both the differences indicate that the epitaxial monolayer graphene is more reactive. The shape of the nanopits is less hexagonal than in graphite and sometimes totally irregular as shown in fig. \ref{f5}(a). However, it should be noted that the present etching time is too short to grow larger and more regularly shaped hexagonal nanopits even in graphite~\cite{MatsuiTBP}. Thus, further studies to optimize the etching conditions are necessary to conclude if better shaped hexagonal nanopits suitable for investigations on the spin-polarized edge states can be fabricated in epitaxial graphene on SiC(0001). We also note that smaller nanopits in the second layer are also present after the H-plasma etching. They preferentially appear near the center of larger nanopits, as shown in fig.~\ref{f4}(b) and fig.~\ref{f5}(a), and expose the SiC substrate. 

Finally, as a brief overview of the preceding discussions, a larger scale STM image and a cartoon of its cross-sectional structure are shown in fig.~\ref{f5}. Figure \ref{f5}(a) shows two substrate regions separated by a SiC step which crosses vertically in the middle of this image. A partly hexagonal nanopit is formed on the left region. Inside the nanopit there is a monolayer QFSG island surrounded by buffer layer, whereas, outside the nanopit, the surface is mostly bilayer QFSG  and partly monolayer graphene on the buffer layer. On the right region, on the other hand, there is a larger, irregularly shaped nanopit with small nanopits in the second layer near the center. In this region, the surface is quasi-free-standing everywhere, consisting of bilayer graphene outside and monolayer graphene inside the nanopits. This cartoon is helpful to understand that at least three different types of graphene zigzag edges were created in this work, i.e., the edges on the H-terminated SiC substrate and on graphene in the quasi-free-standing regions, and the edges on the buffer layer.

\section{Conclusion}
In this study, we have shown the results of morphology and spectroscopy studies of epitaxial graphene on SiC(0001), etched by H-plasma at particular parameters, by STM/S at $T=78$~K. The process produces terraces, which are elevated by $0.12$~nm from the original surface, with irregularly shaped boundaries. The d$I$/d$V$ spectra on such elevated terraces are nearly symmetric with respect to $E_{\mathrm{F}}$ and feature finite density of states with several peaks in the range of $|V_\mathrm{bias}|\leq100$~mV. This is in sharp contrast to the highly asymmetric spectra of the original non-elevated surface. These features indicate that the elevated terrace is floating on the substrate due to inhomogeneous H-intercalation between the buffer layer and SiC(0001). Thus, it was demonstrated that bilayer quasi-free-standing graphene can also be synthesized by H-plasma treatment just as by exposure to atomic hydrogen. In addition, $E_{\mathrm{D}}$ was observed to be relatively close to $E_{\mathrm{F}}$, which makes this sample well suited to investigate the zigzag edge state. We also observed that the etching successfully creates partly hexagonal nanopits with edges aligned to the zigzag direction, on the elevated terraces.

\section*{Acknowledgment}
The authors acknowledge the use of the software WSxM to analyze STM and STS data. A.E.B.A. acknowledges the University of Tokyo Fellowship. This work was financially supported by the Grant-in-Aid for Young Scientists (B) (Grant No. 25800191), Scientific Research (C) (Grant No. 15K05159), and Scientific Research (B) (Grant No. 18H01170) from JSPS.

\bibliographystyle{unsrt}

\end{document}